\documentclass[aps,prc,twocolumn,superscriptaddress,showpacs,floatfix]{revtex4}
\usepackage{graphicx,color}
\usepackage{braket}
\usepackage{amsmath}
\usepackage{amssymb}
\newlength{\figwidth}

\begin{document}
\setlength{\figwidth}{0.98\columnwidth}

\title{Two-proton removal from $^{44}$S and the structure of $^{42}$Si}

\author{J.\,A. Tostevin}
\affiliation{National Superconducting Cyclotron Laboratory, Michigan
State University, East Lansing, MI 48824, U.S.A.}
\affiliation{Department of Physics, Faculty of Engineering and
Physical Sciences, University of Surrey,
Guildford, Surrey GU2 7XH, United Kingdom}
\author{B.\,A.~Brown}
\affiliation{National Superconducting Cyclotron Laboratory, Michigan
State University, East Lansing, MI 48824, U.S.A.}
\affiliation{Department of Physics and Astronomy, Michigan State
University, East Lansing, MI 48824, U.S.A.}
\author{E.\,C. Simpson}
\affiliation{Department of Physics, Faculty of Engineering and
Physical Sciences, University of Surrey,
Guildford, Surrey GU2 7XH, United Kingdom}

\begin{abstract}
Newly-published $^{42}$Si gamma-ray spectra and a final-state-inclusive
$^{42}$Si production cross section value, obtained in a higher-statistics
intermediate-energy two-proton removal experiment from $^{44}$S, are
considered in terms of the final-state-exclusive cross sections computed
using proposed shell-model effective interactions for nuclei near $N=28$.
Specifically, we give cross section predictions when using
the two nucleon amplitudes of the two-proton overlaps $\langle^{42}$Si$
(J^\pi)\,|\,^{44}$S$\rangle$ computed using the newly-proposed {\sc sdpf-mu}
shell-model Hamiltonian. We show that these partial cross sections or their
longitudinal momentum distributions should enable a less-tentative interpretation
of the measured gamma-ray spectra and provide a more quantitative assessment of
proposed shell-model Hamiltonians in this interesting and challenging region
of the chart of nuclides.
\end{abstract}
\date{\today}
\maketitle

\section{Introduction}

Measurements and theoretical predictions of the properties of the
neutron-rich $N=28$ isotones are subjects of considerable interest and
recent studies, in an attempt to understand quantitatively the role of
cross-shell interactions on the erosion of the $N=28$ sub-shell gap
-- particularly for nuclei in the vicinity of $^{42}$Si. This interest,
the major issues involved, and much of the key literature are summarised
concisely in the recent Letter by Takeuchi {\em et al.} \cite{Ta12}, that
also reports a state-of-the-art, higher-statistics, in-flight $\gamma$-ray
spectroscopy measurement of final states of $^{42}$Si. The experiment,
using a fast $^{44}$S($-2$p) two-proton removal reaction, was performed
at the RIBF, RIKEN, at 210 MeV per nucleon. Accessing this region near
$Z=14$ is a considerable experimental challenge. The region also provides
new and severe challenges for theoretical calculations.

The spectrum for $^{42}$Si deduced in Ref. \cite{Ta12} was compared
with the level-scheme from shell-model calculations computed from a
newly-constructed $sd-pf$ cross-shell Hamiltonian, named {\sc sdpf-mu}
\cite{Ut12}. This very recent theoretical study includes and shows
very clearly the importance of the tensor-force component in the
cross-shell $sd-pf$ proton-neutron interaction for reproducing the
measured $2_1^+$- and $4_1^+$-state energies and the $B(E2,0^+
\rightarrow 2^+)$ values in the silicon and sulfur isotopes as $N$
approaches 28 -- including the energy of the (tentatively) assigned
$^{42}$Si $4_1^+$ state in reference \cite{Ta12}.

The analysis of Ref.\ \cite{Ta12} tentatively assigned the observed
1431(11) keV $\gamma$-rays, also seen in a coincidence gate with the
742(8) keV $\gamma$-ray, previously identified as due to the
$2_1^+\rightarrow 0^+$ (g.s.) transition \cite{Ba07}, to the
decay of the $^{42}$Si($4_1^+)$ state located at 2173(14) keV.
The $4_1^+$ state is at 2270 keV according to the {\sc sdpf-mu}
calculation. This and the other {\sc sdpf-mu} bound $^{42}$Si($
J^\pi)$ final states below the (somewhat uncertain) first neutron
threshold, of $3.6(6)$ MeV \cite{NuDat}, are collected in the
first two columns of Table \ref{tbl:gauss}. The additional states
up to 4.2 MeV are also shown. These shell-model
calculations include the full $sd$-shell (for protons) and the
full $pf$-shell (for neutrons) and used the code NuShellX \cite{Nu}.
There are candidate {\sc sdpf-mu} shell-model states for each of
the decay $\gamma$-rays observed in the inclusive spectrum of
\cite{Ta12} since the $4_1^+$, $0_2^+$ and $0_3^+$ shell-model
states all decay through the first excited $2_1^+$ state.

In contrast, a shell-model calculation that uses instead the
{\sc sdpf-u-si} variant of the {\sc sdpf-u} interaction, designed
for $0\hbar\omega$-truncated calculations of the neutron-rich silicon
isotopes in the same model space as is used here \cite{u-si}, produces
a more fragmented spectrum with 23 sub-threshold bound $J^\pi=0^+,
\ldots,\,6^+$, $^{42}$Si final states, 21 of which can be populated
directly by the removal of two $sd$-shell protons. These states, up
to a threshold energy of 3.6 MeV, are shown in Table \ref{tbl:two}.
Two $1^+$ states, at 2985 and 3380 keV, that are found to have
negligible direct 2p-removal cross sections are omitted from the
Table, as are the predicted bound $5^+$ (3288 keV) and $6^+$
(3472 keV) states that cannot be populated directly. There are
a further 9 excited states in the interval from 3.6 to 4.2 MeV.

In the following we calculate the cross sections for population
of the individual $^{42}$Si($J^\pi)$ final states in each case,
resulting from the two-proton removal reaction on a carbon target at
210 MeV per nucleon, the reaction of \cite{Ta12}.

\begin{table}[b]
\caption{Spins, parities and excitation energies of all shell-model
states below the neutron separation energy of $^{42}$Si, 3.6(6) MeV,
computed using the {\sc sdpf-mu} Hamiltonian of Ref.\ \cite{Ut12}.
The additional states up to 4.2 MeV excitation energy are also shown.
The two-proton removal partial cross sections to each final state
and the inclusive cross section for states up to a 3.6 MeV threshold
are also shown, calculated using the {\sc sdpf-mu} TNAs .
\label{tbl:gauss}}
\begin{ruledtabular}
\begin{tabular}{lccc}
$J^\pi$ & $E^{SM}_{\rm SDPF-MU}$ (keV) &$\sigma_{\rm th}(-2$p) (mb) \\
\hline
$0_1^+$ & 0     & 0.186 \\
$2_1^+$ & 821   & 0.031 \\
$4_1^+$ & 2271  & 0.030 \\
$0_2^+$ & 2573  & 0.103 \\
$0_3^+$ & 3273  & 0.011 \\
$2_2^+$ & 3525  & 0.012 \smallskip \\
Inclusive $\sigma_{\rm th}(-2$p):&&   0.37 \smallskip  \\
$2_3^+$ & 3844  & 0.005 \\
$3_1^+$ & 3899  & 0.007 \\
$4_2^+$ & 4080  & 0.010 \\
$2_4^+$ & 4090  & 0.007 \\
\end{tabular}
\end{ruledtabular}
\end{table}

\begin{table}[t]
\caption{Spins, parities and excitation energies of all $^{42}$Si
shell-model states below a 3.6 MeV first neutron threshold energy
having non-negligible direct two-proton removal reaction yields. The
$^{42}$Si states were computed using the {\sc sdpf-u-si} interaction
\cite{u-si}. For the two-proton removal TNA and partial cross section
calculations, the wave function for $^{44}$S was computed using the
{\sc sdpf-u} interaction. \label{tbl:two}}
\begin{ruledtabular}
\begin{tabular}{lccc}
$J^\pi$ & $E^{SM}_{\rm SDPF-U-SI}$ (keV)&$\sigma_{\rm th}(-2$p) (mb) \\
\hline
$0_1^+$ & 0     & 0.249 \\
$2_1^+$ & 815   & 0.025 \\
$0_2^+$ & 1080  & 0.062 \\
$0_3^+$ & 1615  & 0.030 \\
$2_2^+$ & 1622  & 0.015 \\\medskip
$4_1^+$ & 1792  & 0.028 \\
$0_4^+$ & 2398  & 0.015 \\
$3_1^+$ & 2414  & 0.008 \\
$4_2^+$ & 2674  & 0.001 \\
$2_3^+$ & 2709  & 0.013 \\
$2_4^+$ & 2852  & 0.055 \\
$4_3^+$ & 2896  & 0.004 \\
$3_2^+$ & 3033  & 0.039 \\
$2_5^+$ & 3222  & 0.006 \\
$3_3^+$ & 3285  & 0.011 \\
$2_6^+$ & 3400  & 0.002 \\
$4_4^+$ & 3451  & 0.116 \\
$2_7^+$ & 3563  & 0.008 \\\smallskip
$3_4^+$ & 3603  & 0.002 \\
Inclusive $\sigma_{\rm th}(-2$p): &       &   0.69    \\
\end{tabular}
\end{ruledtabular}
\end{table}

\section{Cross section calculations}

The removal of two like nucleons of the deficient species from highly
$N:Z$ asymmetric nuclei has: (a) been shown to proceed directly
\cite{Ba03} and to provide spectroscopic information \cite{To06}, and
(b) has been assessed quantitatively using several 2p- and 2n-removal
reaction test-cases on $sd$-shell nuclei \cite{To06,To04}. We follow the
approach described in detail in these latter references. The structure
details from the shell-model calculations enter the removal reaction
calculations through the two-proton overlaps, i.e.
$\Psi_{JM}^{(f)}(1,2)= \langle1,2,\,^{42}$Si$(JM)
\,|\,^{44}$S$(0^+)\rangle$. Specifically, for each final-state,
$E_x,J$, they generate the set of two nucleon amplitudes (TNAs) $C_{j_1
j_2}^{J}$ for each active two-proton configuration in the assumed
shell-model space where, for a spin $0^+$ projectile,
\begin{eqnarray*}
\Psi_{JM}^{(f)} (1,2) =\sum_{
j_1j_2} C_{j_1j_2}^{J}(-1)^{J+M}/\hat{J}\,
[\overline{\phi_{j_1}(1)\otimes \phi_{j_2}(2)}]_{J-M}\,.
\label{bigone}
\end{eqnarray*}
Given these TNAs, the reaction is described as a sudden, direct removal
process in which the removed protons interact both elastically and
inelastically and reaction residues only elastically with the light
nuclear target, here carbon. These absorptive two-body interactions are
described by their elastic $S$-matrices, computed using the eikonal model
from the complex residue- and proton-target optical potentials. These
potentials, at 210 MeV per nucleon, were computed using the double-
and single-folding models, respectively, assuming a zero-range effective
nucleon-nucleon interaction based on the free pn and pp cross sections.
The point nucleon density of the carbon nuclei was assumed to be of
Gaussian form with a root mean squared (rms) radius of 2.32 fm. That
of the $^{42}$Si residues is discussed below.

As has been detailed for a general case in Ref.\ \cite{Gad08a},
the following nuclear model parameters were used. (a) The point
neutron and proton densities of the $^{42}$Si residues were obtained
from spherical Hartree-Fock calculations using a Skyrme force (the
SkX interaction~\cite{Bro98}). The removed-proton $sd$-shell radial
form factors were described as normalized eigenstates of a Woods-Saxon
plus spin-orbit potential well whose depths were constrained by the
evaluated 2p separation energy $S_{2p} = 40.2(5)$ MeV \cite{Nu} and
the final state excitation energies $E_x$, i.e. $S_p=(S_{2p}+E_x)/2$.
We use excitation energies $E_x=E^{SM}$ as given by the shell-model.
Fixed diffuseness (0.7~fm) and spin-orbit interaction strength (6.0~MeV)
parameters were assumed throughout. The radius parameters of these
Woods-Saxon binding potentials, $r_0$, were determined from a fit to
the rms radius of each proton single-particle orbital of interest,
as given by the HF-SkX calculation. These rms radii play a significant
role in the determination of the removal cross sections, since they
dictate the spatial extent of the two-proton position probabilities
relative to the $^{42}$Si residue densities, and these two sizes must
be specified consistently. Hence our usual practice of using the
same theoretical (HF) model to constrain the proton radii and the
densities used in the evaluation of the interaction $S$-matrices.

The $^{42}$Si residue longitudinal momentum distributions offer
an additional signal of the spins of the populated and $\gamma$-decaying
final-states, as has been discussed formally \cite{ecs1,ecs2} and also
applied for spectroscopy; for example, in identifying the excited
$2_1^+$ and $4_1^+$ states in $^{44}$S following the $^{46}$Ar(-2p)
reaction \cite{SWA11}. In the following section we will discuss
the exclusive momentum distributions of selected {\sc sdpf-mu}
shell-model final states. Those calculations use the same input
parameters as have been detailed above.

\section{Results}
The individual final-state partial cross sections and longitudinal
momentum distributions were calculated following the formalism of
Refs.\  \cite{To06,To04} and \cite{ecs1,ecs2}, respectively, from
the np-form shell-model TNAs computed using NuShellX \cite{Nu}.

\subsection{Exclusive cross sections}
The calculated final-state-exclusive two-proton removal cross
sections for {\sc sdpf-mu} are shown in Table \ref{tbl:gauss},
that shows the strongest populations are of the ground-state
and the first excited $0_2^+$ state. The calculated inclusive
removal cross section is 0.37 mb, to be compared with the reported
value of 0.15(2) mb \cite{Ta12}. So, the ratio of the experimental
to the theoretical yields, $R_s(2N)=\sigma_{\rm exp}/\sigma_{\rm th}
({\rm incl.})=0.41(6)$, is consistent with the systematics, $R_s(2N)
\approx 0.5$, previously observed for a number of $sd$-shell
test-case nuclei \cite{To06}.

The corresponding situation, using the TNAs from the wave functions
for $^{44}$S computed using {\sc sdpf-u} and for $^{42}$Si computed
using {\sc sdpf-u-si}, are shown in Table \ref{tbl:two}. The integrated
cross section to the (21) {\sc sdpf-u-si} final states is 0.69 mb.
The cross section to the lowest 6 states, the $0^+_{1,2,3}$, $2^+_{1,2}$
and $4_1^+$, in common with the {\sc sdpf-mu} bound states (though
reordered), is 0.41 mb. These states all lie below 1.8 MeV when using
{\sc sdpf-u-si}, indicating that significant additional strength is
present at lower excitation energies and that is distributed over a 
large number of the additional bound final states. A very similar integrated
cross section (of 0.62 mb) and pattern of final-state populations arises
if instead one computes both the wave functions for $^{44}$S and $^{42}$Si
using {\sc sdpf-u-si}. These calculated cross-section distributions show
little correspondence with the latest experimental observations \cite{Ta12}.

The predicted relative final-state yields of Table \ref{tbl:gauss}
on the other hand suggest clear fingerprints for a more detailed
assessment of the {\sc sdpf-mu} wave functions from the measured
$\gamma$-ray yields, predicting (i) that 50\% of the inclusive cross
section will pass through the $2_1^+$ state, and (b) specific ratios
for the proposed ($4_1^+$)-state decay strength relative to those for
the $2_1^+$-state and the predicted excited $0_2^+$-state decays.

We note that prediction (i) above is consistent with the observation
from the earlier lower energy and lower statistics $^{44}$S($-2$p)
measurement made at GANIL \cite{Ba07}, which reported that 44(10)\%
of the reaction events were accompanied by a $2_1^+$ to ground state
transition $\gamma$-ray -- there measured to be 770(19) keV in
reasonable agreement with the new data of \cite{Ta12}.

\begin{figure}[t]
\includegraphics[width=0.9\figwidth]{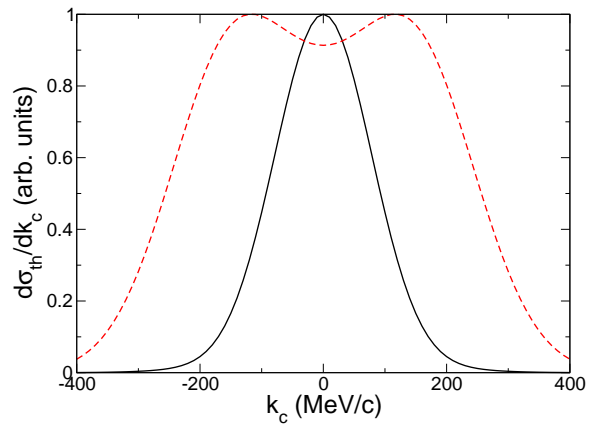}
\caption{(Color online) Calculated exclusive longitudinal momentum
distributions in the projectile rest frame of $^{42}$Si residues
produced in the $0_2^+$ (solid curve) and $4_1^+$ (dashed curve)
shell-model final states of the {\sc sdpf-mu} Hamiltonian \cite{Ut12}.
These calculated momentum distributions are normalized to unity at
their peaks. Their integrated cross sections were shown in Table
\ref{tbl:gauss}. \label{fig1}}
\end{figure}

\subsection{Exclusive momentum distributions}
Further to the fingerprints offered by these relative partial cross
sections, in Figure \ref{fig1} we now show the computed exclusive
longitudinal momentum distributions $d\sigma_{\rm th}(J^\pi)/d k_c$,
in the projectile rest frame, for $^{42}$Si residues produced in
the $4_1^+$ (dashed curve) and $0_2^+$ (solid curve) {\sc sdpf-mu}
shell-model final states. The $4_1^+$ state,
that is dominated by the $(d_{3/2} \otimes d_{5/2})_4$ two-proton
configuration with a dominant $L=4$ total orbital angular momentum
component, is particularly broad. This contrasts markedly with the
$0_2^+$ distribution. These large differential widths suggest that
one could confirm the existence of these states from their momentum
distributions in measurements with quite limited statistics.

\subsection{Inclusive momentum distributions}
\begin{figure}[t]
\includegraphics[width=0.9\figwidth]{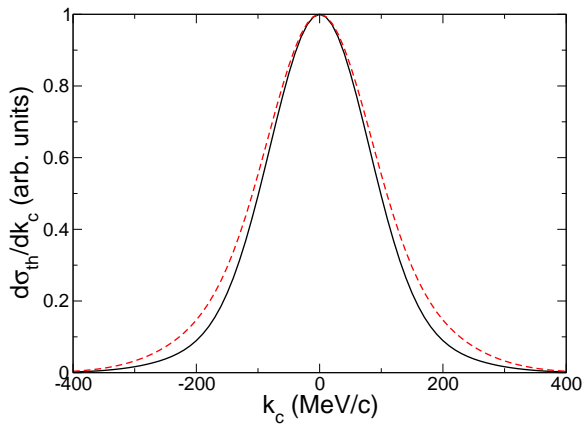}
\caption{(Color online) Calculated inclusive longitudinal momentum
distributions in the projectile rest frame for (i) reactions to all
final states (solid curve) and (ii) for all events with a coincident
$2_1^+$-state $\gamma$-ray (dashed curve). Calculations use the
{\sc sdpf-mu} shell-model TNAs. These momentum distributions are
normalized to unity at their peaks. \label{fig2}}
\end{figure}

Although less specific, the inclusive momentum distributions for
(a) reactions to all final states, and (b) for all events with a
coincident $2_1^+$-state $\gamma$-ray, both involving increased
statistics, would also probe the {\sc sdpf-mu} predictions. The
predicted distributions, obtained by summing the appropriate
exclusive distributions, are shown by the solid and dashed lines,
respectively, in Figure \ref{fig2}. The predicted inclusive cross
section, shown in Table \ref{tbl:gauss} to be dominated by $0^+$
transitions (81\%), is predicted to be narrow, analogous to the
solid, $0_2^+$-state curve shown in Fig.\ \ref{fig1}. The $2_1^+
$-state inclusive events are predicted to be a $60:40$\% admixture
of $0^+:(2^+ + 4^+)$ transitions. This distribution is seen to be
broadened somewhat by the now 40\% component from the $2^+$- and
$4^+$-state distributions. These two distributions thus provide a
valuable, but less specific profile of the predicted {\sc sdpf-mu}
bound states yields, in particular of the $0^+$-states dominance
and, in the case of the coincident $2_1^+$-state events, the relative
importance of excited $0^+$-states of the {\sc sdpf-mu} Hamiltonian.

\section{Summary comments}
We have presented an analysis of the partial cross sections for the
$^{44}$S($-2$p) two-proton removal reaction at 210 MeV per nucleon,
populating the bound final states of $^{42}$Si predicted by the
newly-developed {\sc sdpf-mu} cross-shell Hamiltonian of Ref.\
\cite{Ut12}. The calculations make specific predictions for the
relative populations of the final states, as were shown in Table
\ref{tbl:gauss}, that should be tested against deduced $\gamma
$-ray yields -- as should be available from the data of \cite{Ta12}.

This could allow a considerable strengthening of the tentative
assignment of a 2173(4) keV $4_1^+$ excited state in $^{42}$Si made
in \cite{Ta12} and a more detailed assessment of the predictions of
the {\sc sdpf-mu} Hamiltonian than can provided by, for example,
the $2_1^+$ and $4_1^+$ level energies. This includes the prediction
of a strong population of an excited $0_2^+$ state near 2.5 MeV.
The {\sc sdpf-mu}-based final-state-exclusive reaction yields
presented are consistent with the earlier experimental observation
\cite{Ba07} that 44(10)\% of $^{44}$S($-2$p) reaction events were
accompanied by a $^{42}$Si $2_1^+$ to ground-state decay $\gamma$-ray.

We also show that the computed longitudinal momentum distributions
predict clear additional fingerprints of the populated final-states,
with large differential widths calculated for $^{42}$Si residues
produced in the $4_1^+$ and $0_2^+$ states, states that are predicted
by the {\sc sdpf-mu} Hamiltonian to lie at quite similar excitation
energies. The inclusive momentum distributions are shown to be narrow
reflecting the dominance of $0^+$ final-state transitions predicted
by the {\sc sdpf-mu} Hamiltonian.

In closing we note that complementary $^{42}$Si final-state data
from the $^{43}$P($-$1p) one-proton removal reaction would also
contribute to this $^{42}$Si spectroscopy discussion, since
$sd$-shell proton removal from the $^{43}$P($\frac{1}{2}^+$)
ground-state cannot directly populate proposed $J^\pi=4^+$,
$^{42}$Si final states.

\begin{acknowledgments}
This work was supported by the United Kingdom Science and Technology
Facilities Council (STFC) research grant ST/J000051 and by NSF
grant PHY-1068217.
\end{acknowledgments}

\end{document}